\documentclass[12pt,a4paper]{article}

\usepackage{amsmath}
\usepackage{graphicx}
\usepackage{times}
\usepackage{cite}
\usepackage{ulem}
\begin{document}
\begin{titlepage}

\title{Transition from  the shadow to  reflective scattering}
\author{ S.M. Troshin\footnote{Sergey.Troshin@ihep.ru}, N.E. Tyurin\\[1ex]
\small  \it NRC ``Kurchatov Institute''--IHEP\\
\small  \it Protvino, 142281, Russian Federation}
\normalsize
\date{}
\maketitle

\begin{abstract}
  We consider   transition from shadow to relective scattering mode  and behavior of the amplitude and cross--sections of hadron  interactions in the  both modes. 
\end{abstract}
\end{titlepage}
\setcounter{page}{2}
\section*{Introduction}

 The elastic and inelastic overlap functions introduced by Van Hove \cite{vanh} are related through unitarity with the elastic scattering amplitude (its imaginary part) and  are strongly interdependent.   Transition from a shadow to reflective scattering is determined by an energy evolution of the overlap functions. 
 
  It has been suggested earlier that this evolution will result in reaching the black disc limit  and then its exceeding by the elastic scattering amplitude \cite{plb93}. Such evolution is associated with  formation of a peripheral impact parameter profile of the inelastic overlap function (black ring formation) and is reflecting an increasing effective intensity of soft interactions. It can be  associated \cite{jpg} with formation of a color–conducting medium in the intermediate state of the hadron collision occured at sufficiently high energies and small impact parameters of collision. Similar ideas have been discussed in \cite{geom} and relevant references can be also  found in \cite{lsj}. 
  
  Experimental confirmation of  existence of this effect  has  been found under the LHC experimental data analysis at $\sqrt{s}=13$ TeV (see \cite{tamas}) where  further references  in favor of the black ring formation can be found. This confirmation has an effect beyond $5\sigma$ significance \cite{tamas1} and therefore  the  effect    should not be ignored from the experimental point of view either. Another experimental fact relevant for the discussion of a black ring formation concerns the inreasing ratio of $\sigma_{el}(s)/\sigma_{tot}(s)$. It seems that it indicates an increasing role of confinement in QCD despite that the absolute value of the elastic cross--section does not exceed $1/3$ of the total cross--section magnitude at available energy values. 
  
  Under these studies it should be accounted that the above values are the integrals over the impact parameter, and it is  therefore important to consider the respective differential quantities to reveal their to the unitarity and analiticity constraints.
 
 We discuss the theoretical arguments for  transition from the shadow to reflective scattering mode and use $U$-matrix unitarization scheme (see e.g. \cite{plb93} and references therein) to this end. The above  scheme  provides  a continuious transition between the two scattering modes and is therefore convinient  for a model construction. It does not require    extra  assumption on existence  of the discontinuity  in  energy dependence of the scattering phase   under the standard exponential unitarizaton  \cite{rab}. Indeed, this latter option naturally covers only half of the range allowed by unitarity for the scattering amplitude values.
\section{Standard relations}
A partial elastic scattering matrix element can be represented  as the  complex function:
\begin{equation}\label{sm}
S_l(s)=\kappa_l (s)\exp[2i\delta_l(s)]
\end{equation}
with the two real functions $\kappa_l$ and $\delta_l $.  Herewith $\kappa_l$ can vary in the interval
$0\leq \kappa_l \leq 1$ and is known as an absorption factor connected with  the inelastic overlap function $h_{l, inel}(s)$ (see below).  The value $\kappa_l =0$ means a complete absorption of the respective initial state.
The functions $h_{l, el}(s)$ and $h_{l, inel}(s)$, the contributions of the elastic and inelastic intermediate states into the unitarity relation, are interrelated: 
\begin{equation} \label{unfl}
\mbox{Im} f _l(s)=h_{l,el}(s)+ h_{l,inel}(s).
\end{equation}
The inelastic overlap function can be expressed through the function $\kappa_l(s)$ by  the relation
\begin{equation}
\kappa_l^2 (s)=1-4h_{l,inel}(s).
\end{equation}
The normalization is  $S_l=1+2if_l$ and
 saturation of unitarity relation, $\mbox{Im}f_l\to 1$, leads to $\mbox{Re}f_l\to 0$ and $h_{l,inel}(s)\to 0$ at fixed value of $l$. 
 
 Representation Eq. (\ref{sm}) is able, in principle, to provide  negative values of the real part of the function $S_l$, but an extra assumption on the presence of discontinuity in the $s$ --dependence of the phase $\delta_l(s)$ is required.     Without this assumption, representation of the partial amplitude $f_l(s)$ in the exponential form with the phase $\delta_l(s)$ and factor $\kappa_l(s)$  as the input quantities inevitably limits the consideration by the shadow mode when $|f_l(s)|\leq1/2$. It has just took place in ref. \cite{vanh} where the second solution of the unitarity equation, relevant for the reflective scattering mode $|f_l(s)|>1/2$, was neglected because of  apparent absence of a smooth transition between the shadow and reflective mode.  Indeed, as it was already noted, transition to the reflective mode requires an additional artificial assumption on the critical phase behavior  in the range of energies far beyond the domain populated by resonances.  It would be difficult to reconcile such a behavior with a  physics interpretation of a model based on Eq. (\ref{sm}).
 
 Contrary to that,  the use of the amplitude representation in the rational form ($U$--matrix) and the function $U$ as an input for the model construction  of the hadron scattering amplitude allows one to continuously examine the both scattering modes.    The rational form of unitarization does not require the  singularity presence in the input function, that is an obvious obstacle under model construction. In this sense we use the term ``smooth transition'', and connection of the two scattering modes due to  rational form of the amplitude unitarization naturally serves in favor of the conclusion on the both shadow and reflective modes presence.
 
 Thus, in the framework of $U$-matrix unitarization scheme,  the partial  elastic scattering amplitude $f_l(s)$ is represented as the ratio
 \begin{equation} \label{umfl}
 	 f _l(s)=u_l(s)/[1-iu_l(s)].
 \end{equation}
In the impact parameter representation, Eq. (\ref{umfl}) has the form
  \begin{equation} \label{umfb}
 	f (s,b)=u(s,b)/[1-iu(s,b)].
 \end{equation}
The amplitude $f(s,b)$ is the Fourier--Bessel transform of the scattering amplitude $F(s,t)$:
\begin{equation}\label{imp}
	F(s,t)=\frac{s}{\pi^2}\int_0^\infty bdbf(s,b)J_0(b\sqrt{-t}).
\end{equation}
This unitarization scheme is a solution of unitarity provided that Im$u(s,b)\geq 0$.  

Our aim  is to discuss a continuous transition from the shadow to reflective scattering mode in the   unitarization approach based on  Eqs. (\ref{umfl}) and (\ref{umfb}) and to provide  respective angular distributions  relevant for the both modes.  Such a transition,   if we trace analogy with optics, can be interpreted as the continuous energy increase  of the imaginary part of  a matter refraction index in the transient state.  

As it was noted above, the exponential form,  Eq. (\ref{sm}),  implies a scattering phase discontinuity and therefore is not fully appropriate for  model constructions. However, it should also be noted,  that the exponential form might be modified in order to reproduce  a smooth transition to the negative values of $S$ at the price  of specific   reduction of allowed scattering phase  variation domain and  introduction of an extra free parameter into a scheme \cite{sel} (see also \cite{rab} and \cite{ser} for discussion).

\section{ Angular dependence of the amplitude in the reflective and shadow scattering modes}
Eqs. (\ref{imp}) and (\ref{umfb}) determine the scattering amplitude in the whole region of of transferred momentum variation. The Froissart--Gribov projection formula \cite{fr,gr} assumes an exponential decrease of the amplitude at large values of $b$ and fixed value of the variable $s$. We consider  the pure imaginary case and use the relation
 \begin{equation} \label{uf}
	u (s,b)=f(s,b)/[1-f(s,b)].
\end{equation}

The amplitude $f(s,b)$ is allowed to variate from zero to unity by the unitarity. This is the only restriction imposed on the input function $u$. Actually, the representation of the amplitude in the form of (\ref{umfl}) or (\ref{umfb}) provides the unitarity as well as continuous transition  between the both scattering modes. The variation of $u$ from zero to infinity covers  the both possible intervals of the amplitude variation: the shadow one (0,1/2] and the region  (1/2,1) which can be interpreted as  the reflective scattering range by analogy with the optics \cite{refl}. The reflective  mode appears in the range of variables where the amplitude  values are in the range $1/2 < f(s,b) <1$. It means that  the elastic scattering $S$-matrix element is  in the region $-1<S(s,b) < 0$. The inelastic overlap function
is invariant under sign--changing of $S$ \cite{invar}.

Negative values of $S(s,b)$  correspond to the reflective scattering.
The values of amplitude $f$ in the range $1/2 < f <1$ correspond to $u>1$ and therefore the amplitude cannnot be expanded over the  $u$. 

The  method of the scattering amplitude calculation has been developed long time ago  and is based on the analysis of singularities of the scattering amplitude in the complex $\beta=b^2$ plane \cite{tmf}.   The amplitude expressed in the form of  Eq. (\ref{umfb}) is particularly relevant for such an analysis. The singularities responsible  for the region of small and moderate momentum transfers are the poles in the complex $\beta$--plane. The residues in the poles  $\beta=\beta_n(s)$ allow to obtain the following expression for the scattering amplitude $F(s,t)$ in this region of the  $t$--variation:
\begin{equation}\label{impr}
	F(s,t)\propto \sum_n\sqrt{\beta_n(s)}K_0(\sqrt{t\beta_n(s)}),
\end{equation}
where $K_0(z)$ is the Macdonald function.
Locations of the poles $\beta_n(s)$ are determined by the roots of equation
\begin{equation}\label{pl}
	1+u(s,\beta)=0.
\end{equation}
Those are determined by an explicit form of the function $u$ which should take into account general properties of the scattering matrix under model construction of $u$ encoded into the representation 
\begin{equation}\label{ml}
	u(s,\beta)=\frac{\pi^2}{s}\int_{t_0}^\infty\rho(s,x)K_0(\sqrt{x\beta})dx, 
\end{equation}
 where $\rho(s,x)$ is the spectral density which can be used for the model construction of $u$ and $t_0=4m^2$. In the potential scattering, the function $u$ is proportianal to the integral from the respective potential \cite{bg}:
 \begin{equation}\label{bg}
 	u\propto \int_{-\infty}^\infty V[\sqrt {(z^2+\beta)}]dz .
 \end{equation}
On the base of the Mandelstam representation for elastic scattering amplitude, definition of generalized potential which is complex and energy--dependent was given in \cite{cf}.
 
 The poles $\beta_n(s)$ are  the direct--channel analogues of  the Regge poles. 
 These singularities are generated by  implementation of the unitarity, i.e. by  unitarization of the input amplitude in a rational form while the function $u(s,\beta)$  itself does not have poles at high energies. It has also the branching point at $\beta=0$.  \cite{tmf,tmf1}. 
 
 The   knowledge of the poles  origin is important for  solution of the problem related to separation of the unitarity effects from  effects related to  the adopted input amplitude form and its analytical properties.
 
 There is also separation of contributions related to the different kinds of singularities regarding the specific momentum transfer ranges. 
 In fact, this separation corresponds to  the contributions from different interaction scales.
 We do not consider here the fixed angle scattering region where the amplitude is governed by the singularity other than   the poles, namely by a cut of $u(s,\beta)$ along the negative real axis in the $\beta$-plane \cite{tmf1}.

There is no need to perform continuation into the complex  $b$ or $\beta$ plane for  calculation of the amplitude $F(s,t)$  when $u\leq 1$ since the amplitude $f(s,\beta)$ can be obtained by iteration over  $u$.  
The expansion series  over  $u$ is convergent  and has the form:
\begin{equation}\label{ser}
f(s,\beta)=\sum_{n=1}^\infty (-1)^{n-1}[u(s,\beta)]^n.
\end{equation}
The respective representation for the scattering amplitude $F(s,t)$ is:
\begin{equation}\label{sert}
	F(s,t)=\sum_{n=1}^\infty (-1)^{n-1}U_n(s,t),
\end{equation}
where 
\begin{equation}\label{sertn}
	U_n(s,t)=\frac{s}{2\pi^2}\int_0^\infty d\beta [u(s,\beta)]^nJ_0(\sqrt{-t\beta}).
\end{equation}
Eqs. (\ref{ser})-(\ref{sertn}) correspond to the shadow scattering regime.

The particular case of $u=1$ (at $s = s_r$ and $\beta=0$) is treated by a  limiting transition  $\beta\to 0$. In the impact parameter space the expansions, Eqs. (\ref{impr}) and (\ref{ser}) correspond to the two scattering modes  and reflect  a continuous transition between them.

Regarding the Eq. (\ref{sert}) it should be noted, that there is no clear understanding what the term ``rescattering'' means in the framework of  $S$--matrix formalism operating the asymptotic states. It might happen that the so called successive terms of expansion series correspond, in fact, to  the simultaneous interactions \cite{pesh}. 
\section{Transition from the shadow to reflective scattering}

Both the iteration series Eq.(\ref{sert}) and expansion of the scattering amplitude Eq. (\ref{impr}) result from the chosen unitarization in the rational form.  Unitarity plays a decisive role in the hadron interaction dynamics. 

Another, but interrelated component of this dynamics is associated with a form of the input function $u$.
The particular dependencies of the  $u$ on  energy and impact parameter  are important for  a transition from the shadow to reflective scattering (the value of energy) and, of course, for  other aspects of  hadron interactions. Since the discussion here is  qualitative, we do not make particular model assumptions on  explicit form of the function $u$. We just suppose that this function decreases monotonically with the impact parameter growth in consistence with the  Froissart--Gribov formula for the scattering amplitude \cite{fr,gr}. This formula  is valid at large values of the impact parameter. 

We also suppose that $u(s,\beta)$ monotonically increases with the energy in order to cover the whole range of its allowed values and to provide  saturation of the unitarity limit at $s\to\infty$, (see Eq. (\ref{uf})). This saturation is a necessary prerequisite for  reproduction of the principle of  maximal strength of strong interactions introduced by Chew and Frautchi \cite{chew}. 

The above assumptions imply existence of  the two modes and \it continuous \rm transition to the reflective scattering  starting  at $\beta=0$ when $s=s_r$:
\begin{equation}\label{cdep}
u(s_r,0)=1\implies f(s_r,0)=1/2\;\mbox{and}\;\partial f/\partial u|_{s=s_r, \beta=0}=1/4.
\end{equation}
 The function $u$ is a regular one at   $s=s_r$ contrary to the singular scattering phase $\delta$. Transition to reflective scattering mode generates singularity in the scattering phase \cite{refl}. It makes the  $u$ a preferable quantity for model constructions in the energy region far beyond the resonance region and the use $u$ leads to a continuous dependence of the scattering amplitude $f$ as $s=s_r$, Eq. (\ref{cdep}).

 Genesis of hadron scattering interaction region can be described as a transition from a gray to black disk and then to a black ring with reflective region in the center of this ring.
 The reflective ability  \cite{rab} becomes positive at the energies $s>s_r, \beta=0$.
 Scattering at larger impact parameters remains to be of the shadow nature   moving to the periphery of the interaction region with the energy growth.

The current estimation for the value of $\sqrt{s_r}$  in $pp$--scattering where a  transition from the shadow to reflective scattering starts has been obtained under quantitative analysis of the LHC data \cite{tamas} and has a magnitude $\sqrt{s_r}\simeq 13$ TeV.

\section{Conclusion}
The $U$--matrix representation of the amplitude provides a regular transition between the two scattering modes. This is due to the obvious fact that the amplitude, Eq. (\ref{umfb}), is not limited beforehand by the range of $|f|\leq 1/2$, or the black disk bound.

Transition to the reflective scattering mode would  affect the multiparticle production dynamics \cite{eff}. The related effects, being observed experimentally, would  indicate on  its existence additionally.

The mean multiplicity and average transverse momentum of the produced particles would   slow down  their energy increases at the LHC energies \cite{pl14,exs}.
Expected change of  the multiplicity distribution shape has been described on  base of  dipole Pomeron parameterization of the function $u$ in \cite{str}.

It should be noted again that
asymptotically the total and elastic cross--sections are rising as $\ln^2 s$ while the inelastic cross-section growth is $ \ln s$  at $s\gg s_r$, due to transition to the reflective scattering mode. The difference in  asymptotics of the elastic and inelastic interactions is due to the {\it unitarity limit saturation} by the scattering amplitude and should not depend on a particular way of the amplitude unitarization \cite{rgp}. 

To perform a quantitative analysis of the experimental data,  the explicit models for an input amplitude have to be used.
The most recent illustrations of a quantitative model-based analysis in the framework of $U$--matrix  approach  can be found in refs. \cite{cud,luna}\footnote{We are grateful to V.A. Petrov for draw  our attention to ref. \cite{luna}.}.

As it was noted, appearance of the reflective scattering mode may be interpreted as a formation of a color conducting state of QCD matter  in the transient state at hadron collision \cite {jpg}. Proceeding along this line, one can consider a smooth transition to the reflective scattering mode with energy growth as  a cross--over   between the two phases of matter: the one with hadron degrees of freedom (shadow regime) when the scattering is determined by the multiparticle intermidiate states and the color conducting phase with constituents as the main degrees of freedom (reflective   regime) which  may be associated with high energy head-on collision.


\end{document}